\documentclass[aps,prx,twocolumn,floatfix,longbibliography,superscriptaddress]{revtex4-2}

\usepackage{amsmath}
\usepackage{amssymb}
\usepackage{times}
\usepackage{braket}
\usepackage[pdftex]{graphicx}
\usepackage{color}
\usepackage{blindtext}
\usepackage{nicefrac}
\usepackage[colorlinks=true, citecolor=blue, urlcolor=blue, linkcolor=blue]{hyperref}

\usepackage{todonotes}

\renewcommand{\vec}[1]{\boldsymbol{#1}}

\renewcommand{\ket}[1]{\lvert#1\rangle} 
\newcommand{\braopket}[3]{\langle #1 | #2 | #3\rangle} 

\begin{document}

\title{Orbital Hall effect and orbital altermagnetism in even- and odd-parity-wave magnetic Lieb lattices}

\author{B{\"o}rge G{\"o}bel }
\email[Correspondence email address: ]{boerge.goebel@physik.uni-halle.de}
\affiliation{Institute of Physics and Halle-Berlin-Regensburg Cluster of Excellence CCE, Martin Luther University Halle-Wittenberg, 06120 Halle (Saale), Germany}

\author{Ersoy \c{S}a\c{s}{\i}o\u{g}lu}
\affiliation{Institute of Physics and Halle-Berlin-Regensburg Cluster of Excellence CCE, Martin Luther University Halle-Wittenberg, 06120 Halle (Saale), Germany}

\author{Samir Lounis}
\affiliation{Institute of Physics and Halle-Berlin-Regensburg Cluster of Excellence CCE, Martin Luther University Halle-Wittenberg, 06120 Halle (Saale), Germany}

\date{\today}

\begin{abstract}
Altermagnets combine compensated antiferromagnetic order with ferromagnet-like signatures such as spin-polarized bands and, under appropriate conditions, an anomalous Hall response. Their characteristic momentum-dependent spin splitting originates from the interplay of magnetic order and crystal structure and therefore does not require spin-orbit coupling (SOC), whereas the anomalous Hall effect relies on SOC. This raises the question whether the ferromagnet-like transport character of altermagnets can manifest already in the nonrelativistic limit. Here, we investigate the transport of orbital angular momentum in magnetic Lieb lattices. We show that a collinear antiferromagnetic texture realizes a $d$-wave altermagnetic state and simultaneously generates an orbital Hall effect, both in the complete absence of SOC. In this nonrelativistic limit, the orbital Hall response closely resembles that of the corresponding ferromagnet. When SOC is included, the orbital Hall effect is accompanied by a spin Hall response, while the altermagnetic spin texture acquires a corresponding orbital texture, realizing orbital altermagnetism. In contrast, an anomalous, or crystal, Hall effect requires SOC and is additionally subject to crystal-symmetry constraints. We extend the analysis from $d$-wave altermagnetism to odd-parity $p$-wave magnetism, where the orbital Hall effect becomes anisotropic and the orbital conductivity tensor develops a symmetric transverse component analogous to the planar Hall response in charge transport. Our results establish the orbital Hall effect as a nonrelativistic transport manifestation of the close relation between altermagnets and ferromagnets and extend this connection to unconventional magnetic orders beyond altermagnetism.
\end{abstract}

\maketitle



\noindent Altermagnets~\cite{hayami2019momentum,vsmejkal2020crystal,hayami2020bottom,ma2021multifunctional,yuan2021prediction,smejkal2022beyond,krempasky2024altermagnetic,amin2024nanoscale,fedchenko2024observation} combine characteristic properties of ferromagnets and conventional antiferromagnets. Their magnetic moments compensate within the unit cell, resulting in a vanishing net magnetization, while their electronic bands exhibit a momentum-dependent spin splitting reminiscent of a ferromagnet. Unlike conventional spin splitting induced by spin-orbit coupling (SOC), the altermagnetic spin splitting originates from the combination of magnetic order and crystal symmetry and therefore persists in the nonrelativistic limit. This peculiar coexistence of compensated magnetic order and spin-polarized electronic states has established altermagnets as a distinct class of magnetic materials with promising implications for spintronics~\cite{vsmejkal2022emerging,bai2024altermagnetism,song2025altermagnets,tamang2025altermagnetism,jungwirth2025altermagnetism}.

A particularly intriguing aspect of altermagnetism is its connection to transverse transport. Despite their compensated magnetization, altermagnets can exhibit an anomalous Hall effect, often referred to as the crystal Hall effect in this context~\cite{vsmejkal2020crystal}, since its existence and sign are governed by the interplay of magnetic order and crystal symmetry. However, there is an important distinction between this response and the characteristic altermagnetic spin splitting: While the latter exists without SOC, the crystal Hall effect generally relies on SOC and can additionally be prohibited by crystal symmetries. The often invoked analogy between altermagnets and ferromagnets in terms of the anomalous Hall effect therefore relies not only on a relativistic interaction that is not required for altermagnetism itself, but also on additional symmetry requirements. This is why we search for a different transport phenomenon that more generally reflects the ferromagnet-like character of an altermagnet already in the absence of SOC.

\begin{figure}[t!]
    \centering
    \includegraphics[width=0.9\columnwidth]{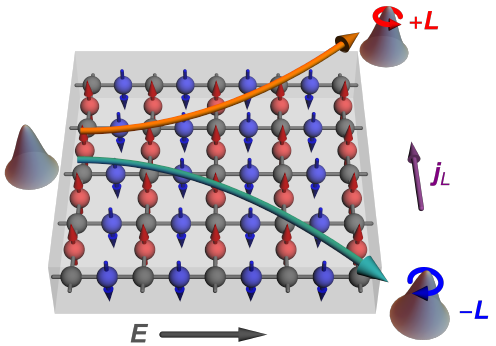}
    \caption{\textbf{Orbital Hall effect in a Lieb lattice altermagnet.} In the presence of an electric field $\vec{E}$, an electronic wave packet propagates through the Lieb lattice altermagnet that exhibits a collinear antiferromagnetic texture. As a consequence, states with positive and negative orbital angular momentum $\vec{L}$ get deflected into opposite transverse directions. The result is a transverse orbital current $\vec{j}_{\vec{L}}$ corresponding to an orbital Hall effect.}
    \label{fig:overview}
\end{figure}

\begin{figure*}[t!]
    \centering
    \includegraphics[width=0.8\textwidth]{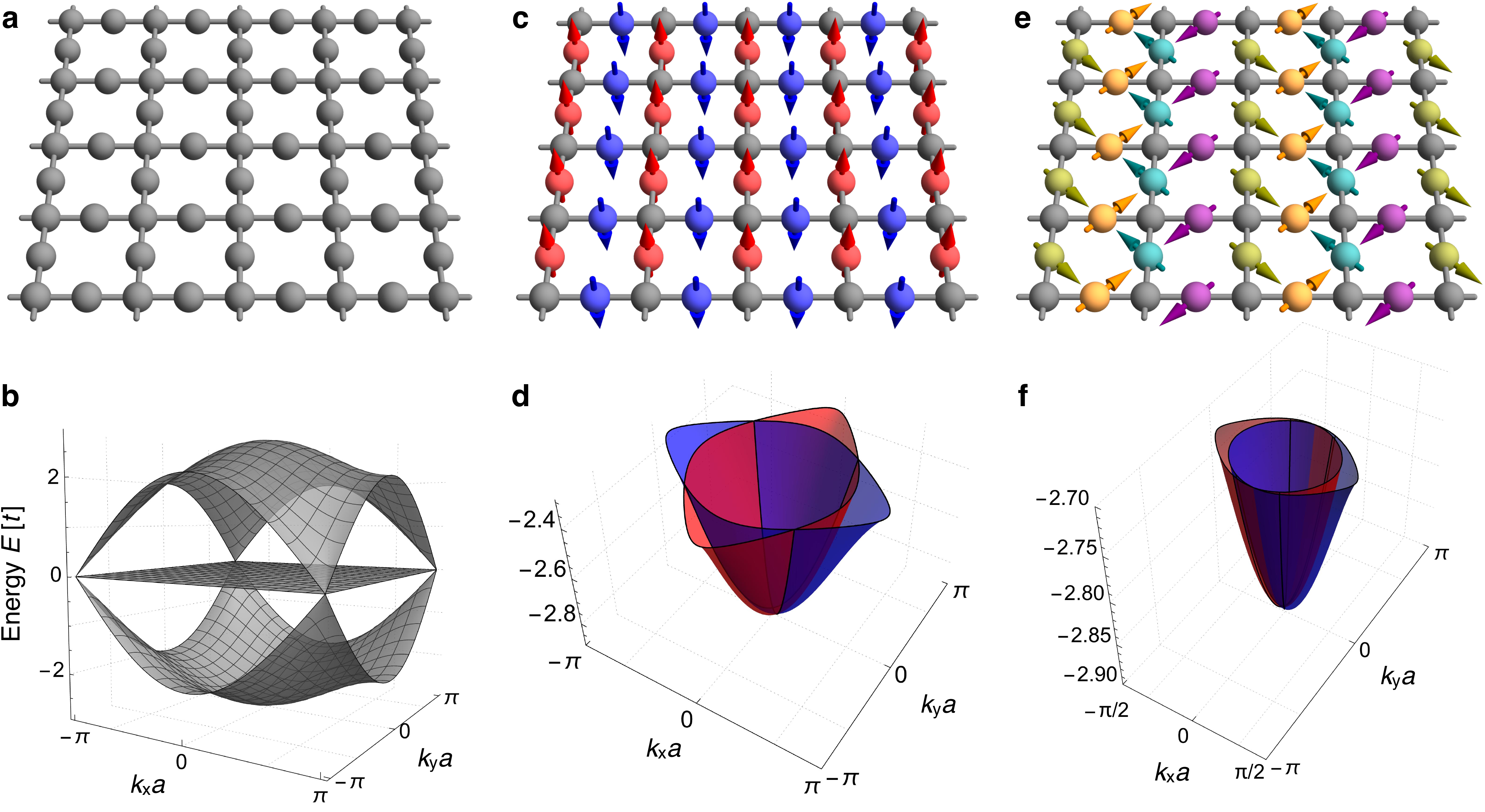}
    \caption{\textbf{Systems investigated in this paper.} \textbf{a} Nonmagnetic Lieb lattice, and \textbf{b} corresponding band structure. \textbf{c} Collinear antiferromagnetic texture on a Lieb lattice turning it into an altermagnet. The arrows represent the magnetic moments. \textbf{d} Corresponding band structure with anisotropic spin splitting. Red: spin up, blue: spin down. The full band structure is shown in Fig.~\ref{fig:altermagnet}. \textbf{e} Spin spiral on a Lieb lattice turning it into a p-wave magnet. \textbf{f} Corresponding band structure. The full band structure is shown in Fig.~\ref{fig:antialtermagnet}.}
    \label{fig:systems}
\end{figure*}

Naturally, the orbital degree of freedom~\cite{zhang2005intrinsic, bernevig2005orbitronics, kontani2008giant, tanaka2008intrinsic, kontani2009giant,go2018intrinsic,salemi2022theory,busch2023orbital,choi2023observation,lyalin2023magneto,gobel2024OHE,gobel2024topological,yoda2015current,go2017toward,el2023observation,gobel2025chirality} comes to mind. Orbital angular momentum~\cite{chang1996berry,xiao2005berry,thonhauser2005orbital,ceresoli2006orbital,raoux2015orbital,gobel2018magnetoelectric,pezo2022orbital} is intimately connected to the motion of electrons through the crystal and can therefore couple directly to its structure even in the absence of SOC. This makes orbital phenomena particularly appealing in the context of altermagnetism, where the crystal symmetry itself is essential for generating the characteristic spin splitting. 
A prominent example for orbital transport is the orbital Hall effect~\cite{zhang2005intrinsic, bernevig2005orbitronics, kontani2008giant, tanaka2008intrinsic, kontani2009giant,go2018intrinsic,salemi2022theory,busch2023orbital,choi2023observation,lyalin2023magneto,gobel2024OHE,gobel2024topological} (shown in Fig.~\ref{fig:overview}), where an applied electric field generates a transverse flow of orbital angular momentum without requiring SOC. In fact, orbital Hall transport has recently been theoretically investigated in the altermagnet RuO$_2$~\cite{yahagi2024neel,zhang2026coexistence}. Both altermagnetism and orbital Hall transport can originate from the crystal structure without relying on relativistic interactions, suggesting an intrinsic connection between the two. The orbital Hall effect therefore provides a natural candidate for a transport phenomenon that connects altermagnets and ferromagnets already in the nonrelativistic limit.

\begin{figure*}[t!]
    \centering
    \includegraphics[width=0.8\textwidth]{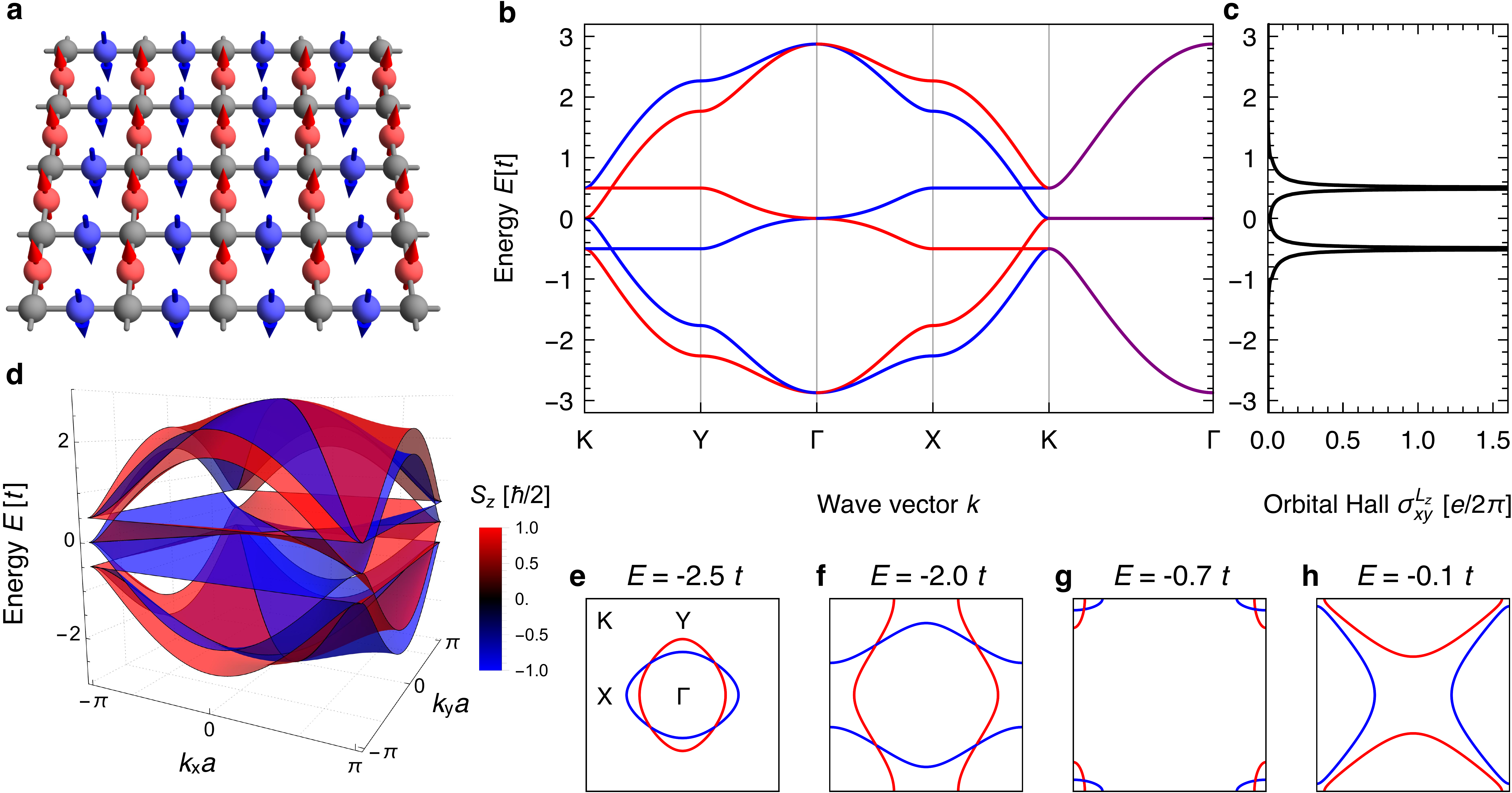}
    \caption{\textbf{Altermagnetic Lieb lattice without spin-orbit coupling.} \textbf{a} Lieb lattice with collinear antiferromagnetic spin texture realizing an altermagnet. \textbf{b} Band structure along high symmetry directions (red: positive $S_z$, blue: negative $S_z$). \textbf{c} Orbital Hall conductivity. \textbf{d} Band structure in the full Brillouin zone. \textbf{e-h} Fermi lines at various energies.}
    \label{fig:altermagnet}
\end{figure*}

Here, we demonstrate a connection between altermagnetism and orbital Hall transport using a minimal tight-binding model on the Lieb lattice (Fig.~\ref{fig:systems}a,b). A collinear antiferromagnetic texture with compensated magnetization (Fig.~\ref{fig:systems}c) generates the characteristic anisotropic spin splitting of a d-wave altermagnet (Fig.~\ref{fig:systems}d) and, at the same time, a finite orbital Hall effect -- both without SOC (Fig.~\ref{fig:altermagnet}). The simplicity of the model allows us to directly compare nonmagnetic, conventional antiferromagnetic, altermagnetic, and ferromagnetic configurations. We find that the altermagnet closely resembles the ferromagnet in its orbital transport, whereas the corresponding conventional antiferromagnet exhibits no orbital Hall effect. The orbital Hall effect thus provides a nonrelativistic transport manifestation of the close relation between altermagnets and ferromagnets.

\begin{figure*}[t!]
    \centering
    \includegraphics[width=.9\textwidth]{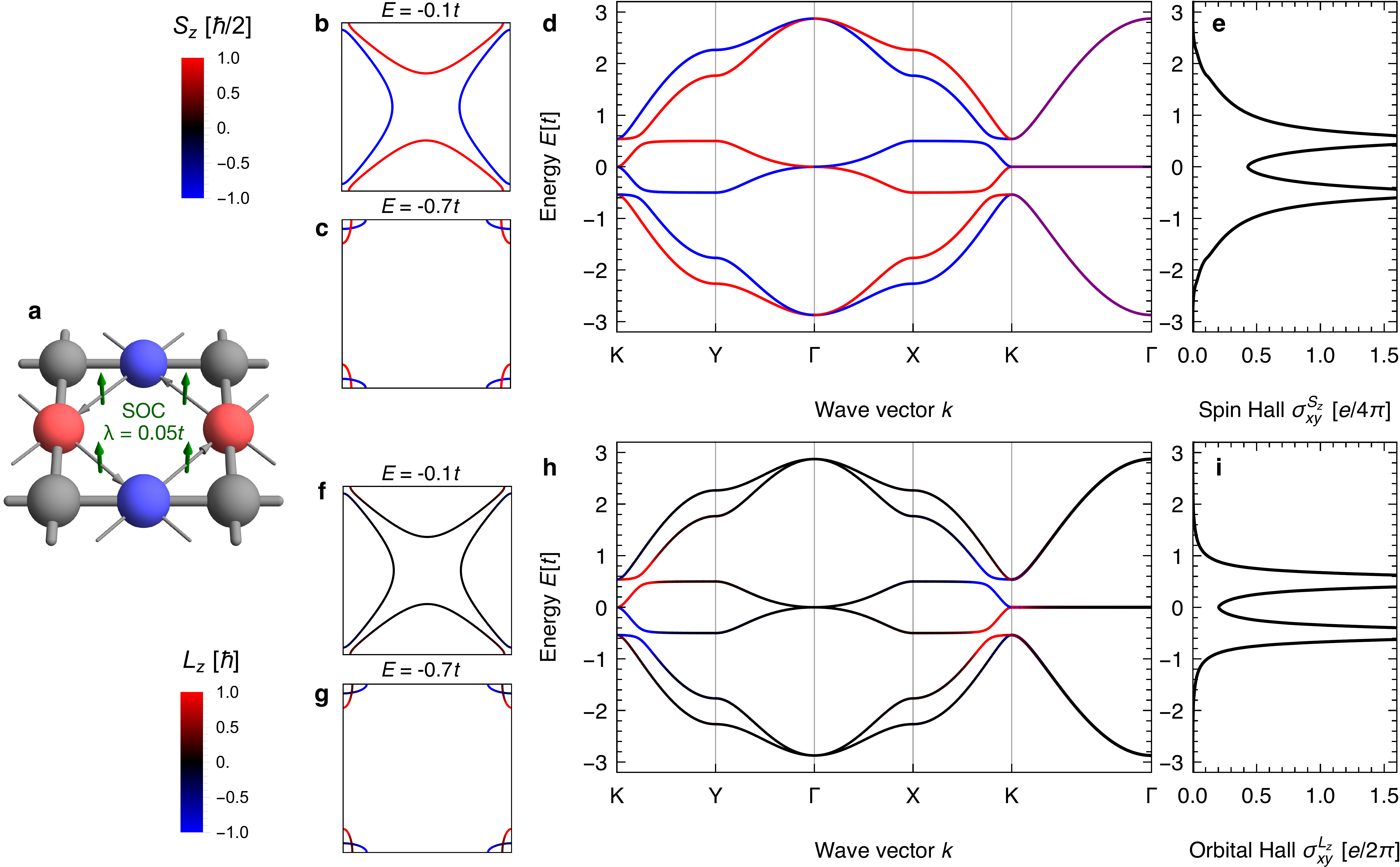}
    \caption{\textbf{Altermagnetic Lieb lattice with spin-orbit coupling.} \textbf{a} Lieb lattice with spin-orbit coupling entering via second-nearest neighbor hoppings. The green arrows represent the chirality of the hopping path determining the sign of $\nu_{ij}=\pm1$ entering the Hamiltonian. \textbf{b,c} Fermi lines (red: positive $S_z$, blue: negative $S_z$). \textbf{d} Spin-polarized band structure. \textbf{e} Spin Hall conductivity. \textbf{f,g} Same Fermi lines with red: positive $L_z$, blue: negative $L_z$. \textbf{h} Orbital-polarized band structure. \textbf{i} Orbital Hall conductivity.}
    \label{fig:soc}
\end{figure*}

When SOC is introduced, an intriguing correspondence between spin and orbital phenomena emerges. The orbital Hall effect is accompanied by a spin Hall response, while the characteristic altermagnetic spin texture acquires a corresponding orbital texture in reciprocal space (Fig.~\ref{fig:soc}). In this sense, SOC converts the system into an orbital altermagnet. Additional crystal-symmetry breaking allows for an anomalous Hall response, realizing the crystal Hall effect. 

\begin{figure*}[t!]
    \centering
    \includegraphics[width=.85\textwidth]{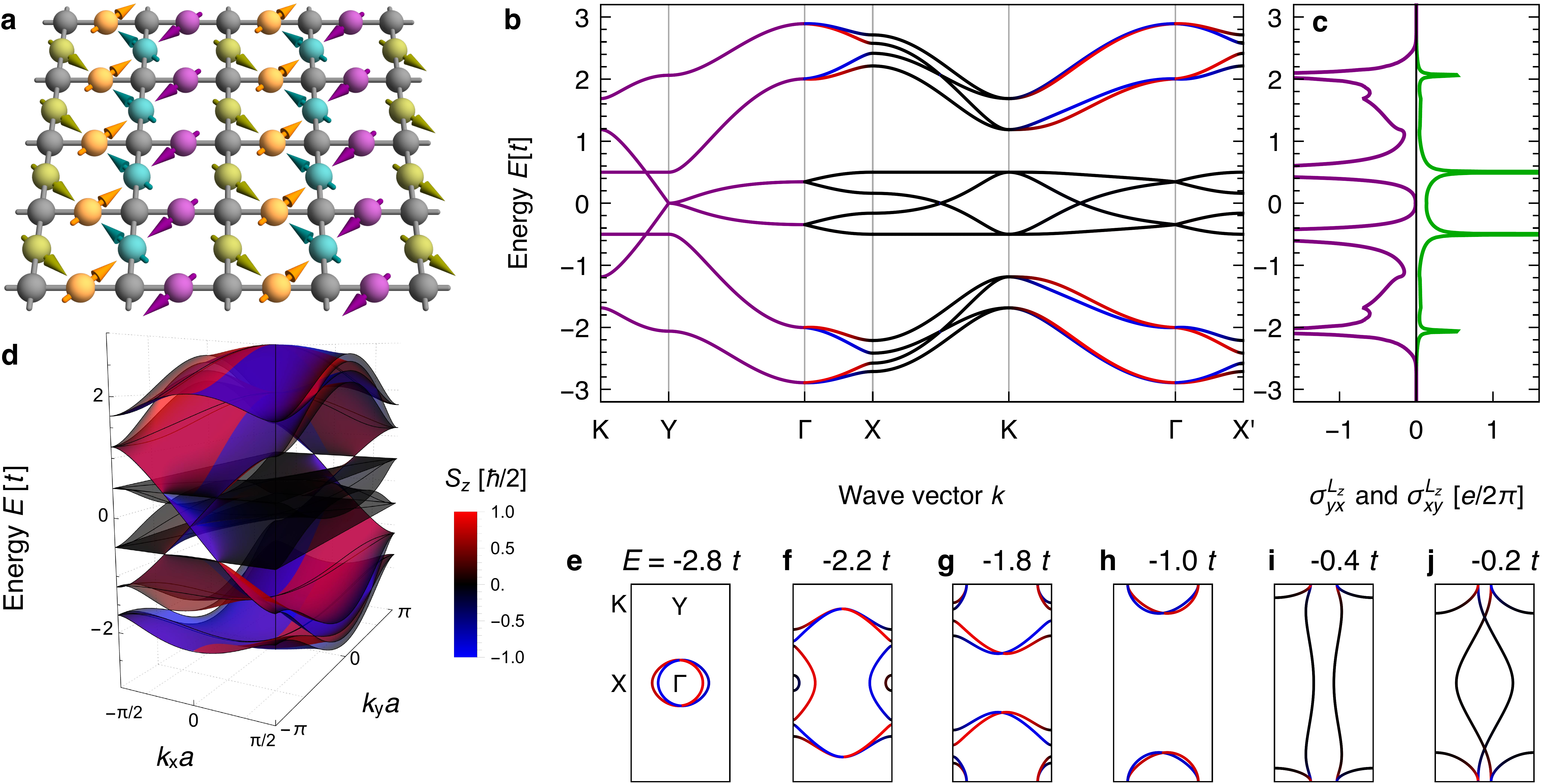}
    \caption{\textbf{p-wave magnetic state of the Lieb lattice.} \textbf{a} Lieb lattice with spin-spiral texture. \textbf{b} Band structure along high symmetry directions (red: positive $S_z$, blue: negative $S_z$). \textbf{c} Orbital Hall conductivity (purple: $\sigma_{yx}^{L_z}$, green: $\sigma_{xy}^{L_z}$). \textbf{d} Band structure in the full Brillouin zone. \textbf{e-j} Fermi lines at various energies.}
    \label{fig:antialtermagnet}
\end{figure*}

Finally, we extend our analysis to an odd-parity-wave magnet~\cite{hellenes2023p}, i.\,e., a spin spiral on a Lieb lattice (Fig.~\ref{fig:systems}e), characterized by p-wave rather than d-wave spin splitting (Fig.~\ref{fig:systems}f). We find that it likewise supports an orbital Hall effect without SOC, with the anisotropy of the p-wave magnetic state directly reflected in a non-antisymmetric orbital Hall conductivity tensor (Fig.~\ref{fig:antialtermagnet}).

Our results establish orbital transport as a nonrelativistic link between the seemingly antiferromagnetic and ferromagnetic characteristics of altermagnets and extend this connection to unconventional magnetic orders beyond altermagnetism.\\
%
%
%
\\
\noindent\textbf{Results and Discussion}\\
\noindent\textbf{Orbital Hall effect in altermagnetic Lieb lattices}


\noindent In this paper, we consider a two-dimensional (inverse) Lieb lattice~\cite{lieb1989two} as shown in Fig.~\ref{fig:systems}. If not stated otherwise, we consider nearest-neighbor hopping with amplitude $t$. The three basis atoms can host various magnetic textures $\{\vec{m}_i\}_i$ and might differ in onsite energies $o_i$. This system allows us to construct a minimal model of a d-wave altermagnet~\cite{brekke2023two,chang2026inverse}, as pointed out below, and is realized in materials such as FeSe~\cite{mazin2023induced}, La$_2$O$_3$Mn$_2$Se$_2$~\cite{wei20252}, V$_2$O~\cite{sasioglu2026chiral} as well as V$_2$Te$_2$O~\cite{jaeschke2025atomic} and V$_2$Se$_2$O~\cite{parthenios2025spin}.

The tight-binding Hamiltonian is
\begin{align} 
  H &  =  -t \sum_{\braket{ij}} \,c_{i}^\dagger c_{j} + m \sum_{i} \vec{m}_{i} \cdot (c_{i}^\dagger \boldsymbol{\sigma}c_{i})+\sum_i o_i\,c_{i}^\dagger c_{i}.
  \label{eq:ham} 
\end{align} 
In second quantization, $c_{i}^\dagger$ and $c_{i}$ are the creation and annihilation operators of an electron at site $i$ and $\vec{\sigma}$ is the vector of Pauli matrices. The first term is the hopping and the second term is the Hund's coupling between conduction electrons and the magnetic texture that is formed by lower-lying electronic states that are not explicitly considered. The third term describes the onsite energies. SOC will be added in form of a Kane-Mele term in a later section of this paper.

By diagonalizing this Hamiltonian, we determine the band structure $E_{\nu\vec{k}}$ and eigenvectors $\ket{\nu\vec{k}}$ with band index $\nu$ and wave vector $\vec{k}$. From these, we can calculate the spin texture $\vec{S}_{\nu}(\vec{k})$, orbital texture $\vec{L}_{\nu}(\vec{k})$ and the anomalous Hall conductivity $\sigma_{xy}$, spin Hall conductivity $\sigma_{xy}^{S_z}$ and orbital Hall conductivity $\sigma_{xy}^{L_z}$, as described in the Methods section. To capture the intersite-contributions to the orbital angular momentum, we consider the modern theory of orbital magnetization ~\cite{chang1996berry,xiao2005berry,thonhauser2005orbital,ceresoli2006orbital,raoux2015orbital,gobel2018magnetoelectric,pezo2022orbital} that goes beyond the atomic-center approximation and is crucial for capturing geometric contributions to the orbital Hall effect~\cite{busch2023orbital}.

The non-magnetic Lieb lattice ($m=0$), as shown in Fig.~\ref{fig:systems}a, gives rise to a spin-degenerate band structure of 3 band pairs (see Fig.~\ref{fig:systems}b). The bands are neither spin- nor orbital-polarized and there is no Hall transport of charge, spin and orbital angular momentum (see Supplementary Fig.~S1 for a more detailed analysis). The middle band pair is perfectly flat which is the characteristic signature of the Lieb lattice. It corresponds to a localized mode that does not propagate.

This changes once we consider a collinear antiferromagnetic texture. We use $m=0.5t$ and $\vec{m}_{1,2}=\mp\vec{e}_z$ and $\vec{m}_3=0$, corresponding to Fig.~\ref{fig:systems}c. As initially proposed in Ref.~\cite{brekke2023two}, the non-magnetic atom makes the system altermagnetic: The band structure is spin-polarized with an anisotropic spin splitting even though the net magnetization is compensated (see Fig.~\ref{fig:systems}d).

The full band structure is shown in Fig.~\ref{fig:altermagnet}b,d with Fermi surfaces in panels e-h. The bands are only spin degenerate along the high-symmetry direction $\Gamma \mathrm{K}$ but are spin polarized elsewhere. Despite the anisotropic spin splitting, the altermagnetic Lieb lattice does not give rise to a spin Hall or anomalous Hall effect since SOC is not considered yet. However, the system exhibits an orbital Hall effect (Fig.~\ref{fig:altermagnet}c) even though there is no orbital texture $L_{z,\nu}(\vec{k})=0$. The orbital Hall conductivity $\sigma_{xy}^{L_z}$ is always positive and diverges at the energies $\pm m$ which is where the formerly flat band pair intersects the formerly dispersive bands. This means the formerly localized mode couples to the propagating modes, thereby allowing for transport of orbital angular momentum.

The above described properties of the altermagnet are akin to those of a ferromagnet (see Fig.~S2 in the Supplementary Material). Without SOC, the ferromagnet exhibits spin-polarized bands but does not cause an anomalous or spin Hall effect. However, as our calculations reveal, there is a non-zero orbital Hall effect in the ferromagnetic Lieb lattice. 
In this sense, our calculations confirm the statement that an altermagnet is an antiferromagnet based on the magnetic texture but behaves like a ferromagnet based on the transport. 
Since altermagnetism is defined strictly without SOC, it is highly significant that we found a common transport signature of ferromagnets and altermagnets that emerges without SOC: the orbital Hall effect.

When we consider a stronger coupling to the antiferromagnetic texture $m>t$ (see Fig.~S3 in the Supplementary Material), we still see the characteristic altermagnetic spin splitting. The band structure now exhibits 3 clearly separated band pairs. The middle band pair, centered around $E=0$, corresponds to the non-magnetic atom. Since there are no intersections of the band pairs, there are no divergences anymore and the system becomes an orbital Hall insulator with orbital Hall plateaus in the band gap. Overall, the orbital Hall conductivity is much smaller because the system behaves more similarly to a conventional antiferromagnet: Results for a checkerboard antiferromagnet, for which the non-magnetic atom has been removed and hopping between the magnetic atoms has been considered, are shown in Fig.~S4 in the Supplementary Material. All bands are Kramers-degenerate and there is no orbital Hall effect.\\
%
%
%
%
%
%
\\
\noindent\textbf{Orbital altermagnetism and anomalous Hall effect caused by spin-orbit coupling}\\
\noindent Next, we consider SOC similarly to the Kane-Mele model~\cite{kane2005z}. We add a spin-dependent second-nearest neighbor hopping term to the Hamiltonian 
\begin{align}
    H_\mathrm{SOC}=i\lambda\sum_{\braket{\braket{i,j}}}\nu_{ij}c_i^{\dagger}\sigma_z c_j.
\end{align}
The sign $\nu_{ij}=\pm1$ depends on the curvature of the nearest neighbor hopping path towards the second-nearest neighbors. An example is shown by the green arrows in Fig.~\ref{fig:soc}a. Upwards-pointing arrows resemble $+1$ and downwards-pointing arrows $-1$ for the respective path. In the following, we use a strength of $\lambda=0.05t$.

The top half of Fig.~\ref{fig:soc} shows the spin-dependent properties. Panels b-d are the Fermi lines and band structure, respectively. In comparison to the results without SOC (cf. Fig.~\ref{fig:altermagnet}), the only qualitative change is that the crossings between the formerly flat and dispersive band pairs open. This is also the reason, why the orbital Hall conductivity in panel (i) does not diverge anymore. Due to the coupling of spin and orbital angular momentum, a spin Hall conductivity emerges as well, as shown in panel e. Both Hall conductivities exhibit similar energy dependencies. Here we have used positive $\lambda$. For negative $\lambda$, only the spin Hall conductivity changes sign. As in most systems, the orbital Hall conductivity arises without SOC and additionally converts into a spin Hall conductivity once SOC is considered. 

The opposite causality occurs in terms of the magnetic texture in reciprocal space: The anisotropic spin splitting, that is present without SOC, converts into an anisotropic orbital texture once SOC is considered. So in that sense, the Lieb lattice becomes also a $d$-wave orbital altermagnet; however, only in the presence of SOC: The orbital angular momentum texture fulfills $\vec{L}(\vec{k})=-\vec{L}(-\vec{k})$ as in non-magnetic chiral crystals like Tellurium~\cite{gobel2025chirality,oh2026observation} or in chiral carbon nanotubes~\cite{gobel2025chirality2}.

However, the altermagnetic Lieb lattice does not exhibit an anomalous Hall effect, even once SOC is considered. The reason is that a combined $C_4\mathcal{T}$ symmetry of time-reversal $\mathcal{T}$ and $90^\circ$ rotation of the lattice around the $z$ axis $C_4$ forces the anomalous Hall conductivity to vanish. This illustrates that SOC alone is not sufficient to generate an anomalous Hall effect in an altermagnet, which additionally has to be allowed by the crystal symmetry. To demonstrate this distinction, we explicitly break the $C_4\mathcal{T}$ symmetry by varying the on-site energies. 

In Fig.~S5 in the Supplementary Material, we compare the altermagnetic system with $o_1=0.2t$, $o_{2,3}=0$ with a ferromagnetic system. Both systems exhibit anomalous, spin and orbital Hall effects once SOC is considered. In contrast, the orbital Hall effect requires neither SOC nor this artificial crystal-symmetry breaking and is already present in the pristine altermagnetic Lieb lattice.

In Fig.~S6 in the Supplementary Material, we consider the non-magnetic and antiferromagnetic Lieb lattice for comparison. In both of these systems the anomalous Hall effect is absent, even in the presence of SOC and even if the onsite energies are varied. This shows that the altermagnet is in fact more similar to the ferromagnet than to an antiferromagnet or non-magnet in terms of transport properties.

Lastly, we have analyzed the angle dependence of the transport properties on the Néel vector orientation in the altermagnetic system. The results are shown in Fig.~S7 in the Supplementary Material. Without SOC, the Hall transport properties do not depend on the Néel vector orientation at all. However, once SOC is considered, the Hall responses depend on the out-of-plane angle $\theta$. Data for $\theta=\pm 90^\circ$, $\pm45^\circ$ and $0^\circ$ are shown as examples. The angle dependence can be explained by the fact that spin-orbit interaction is proportional to $\vec{S}\cdot\vec{L}$ and $\vec{L}$ only has an out-of-plane component in our system.\\
%
%
%
%
%
%
\\
\noindent\textbf{Anisotropic orbital Hall effect in p-wave magnet}\\
\noindent Unlike the intrinsic anomalous Hall conductivity, the orbital conductivity is not generally required to be antisymmetric, such that $\sigma_{xy}^{L_z}\neq-\sigma_{yx}^{L_z}$ is allowed. In the altermagnetic Lieb lattice, the above mentioned $C_4\mathcal{T}$ combined four-fold rotation around $z$ and time reversal symmetry relates the $x$ and $y$ directions and enforces $\sigma_{xy}^{L_z}=-\sigma_{yx}^{L_z}$. However, once this symmetry is broken, the $x$ and $y$ directions become inequivalent. Consequently, $\sigma_{xy}^{L_z}$ and $\sigma_{yx}^{L_z}$ can become independent and indeed, we already find a small symmetric component of the orbital conductivity in the case of an altermagnetic configuration with a difference in onsite energies, discussed before (see Fig.~S5d).

This motivates us to investigate spin-spirals on a Lieb lattice that break the $C_4\mathcal{T}$ symmetry and are known to give rise to odd-parity-wave magnetism~\cite{hellenes2023p}. Instead of exhibiting a parity-even d-wave spin splitting, the state exhibits an odd p-wave spin splitting. Such antisymmetric spin splitting $\vec{s}(\vec{k})=-\vec{s}(-\vec{k})$ based on spin spirals~\cite{brekke2024minimal,gobel2026exact} has for example been realized in systems such as CeNiAsO~\cite{hellenes2023p}, NiI$_2$~\cite{song2025electrical} and Gd$_3$(Ru,Rh)$_4$Al$_{12}$~\cite{yamada2025metallic}. 

On the Lieb lattice, we can generate a p-wave magnet by considering a twice as large unit cell as before and using an in-plane spin spiral as the magnetic texture; cf. Fig.~\ref{fig:antialtermagnet}a. The band structure (panels b,d) exhibits the characteristic spin splitting that also becomes visible in the Fermi lines (panels e-j). The p-wave magnet gives rise to an orbital Hall effect (panel c). We observe an anisotropy of the band structure comparing $k_x$ and $k_y$ and an anisotropy in the orbital conductivity tensor, i.\,e., $\sigma_{xy}^{L_z}\neq-\sigma_{yx}^{L_z}$, as discussed above. The resulting orbital conductivity can be decomposed into an antisymmetric Hall-like contribution $\sigma^{L_z}_A=(\sigma^{L_z}_{xy}-\sigma^{L_z}_{yx})/2$ and a symmetric transverse contribution $\sigma^{L_z}_S=(\sigma^{L_z}_{xy}+\sigma^{L_z}_{yx})/2$, the latter being analogous to the symmetric off-diagonal response underlying the planar Hall effect~\cite{goldberg1954new} in charge transport.

Note that the $p$-wave configuration does not exhibit an anomalous Hall effect even though $C_4\mathcal{T}$ is broken. This is due to a remaining $\tau_{1/2}\mathcal{T}$ symmetry, combining time reversal with a translation by half the magnetic unit cell $\tau_{1/2}$, which enforces a vanishing anomalous Hall conductivity. In contrast, this symmetry does not prohibit an orbital Hall response and, in particular, does not impose $\sigma_{xy}^{L_z}=-\sigma_{yx}^{L_z}$.

When we reverse the chirality of the spin spiral, the sign of the spin reverses, so the spin polarization of the p-wave-split bands is reversed. However, the orbital Hall conductivity is chirality insensitive. This is because the two spirals of opposite chirality are related by a global rotation of $\pi$ in spin space around an in-plane axis. In the absence of SOC, this transformation reverses the relevant spin texture $S_z(\vec{k})$ while leaving purely orbital observables invariant.

Once SOC is considered (see Fig.~S8), a spin Hall effect emerges as well. The spin conductivity tensor is also not perfectly antisymmetric but the symmetric part $\sigma^{S_z}_S=(\sigma^{S_z}_{xy}+\sigma^{S_z}_{yx})/2$ is much smaller. The reason for this can be attributed to the fact that the spin is an intrinsic property of the conduction electrons while the orbital angular momentum is tied to the lattice degree of freedom and is affected by the anisotropy of the system along $x$ and $y$ more strongly. Like for the altermagnetic configuration, we also see here that the characteristic spin splitting (here p-wave) converts to a p-wave orbital splitting; i.\,e., we have realized an orbital p-wave magnet but only in the presence of SOC. This latter characteristic holds for spin spirals in general, not just on the Lieb lattice~\cite{saunderson2026coupled}.\\
%
%
\\
\noindent\textbf{Conclusion}\\
\noindent In summary, we have investigated the interplay between unconventional magnetic order and orbital transport in a minimal tight-binding model on the Lieb lattice. We find that the collinear compensated magnetic texture generating the characteristic altermagnetic spin splitting simultaneously gives rise to an orbital Hall effect without SOC. Comparing different magnetic configurations of the same model reveals a close correspondence between the altermagnet and ferromagnet: both exhibit an orbital Hall effect in the nonrelativistic limit, whereas it is absent in the corresponding conventional antiferromagnet. This establishes the orbital Hall effect as a nonrelativistic transport manifestation of the often invoked relation between altermagnets and ferromagnets.

SOC further intertwines the spin and orbital degrees of freedom. The orbital Hall effect is then accompanied by a spin Hall response, while the altermagnetic spin texture acquires a corresponding orbital texture, rendering the system an orbital altermagnet. An anomalous Hall effect in the form of a crystal Hall effect can additionally emerge if the crystal symmetries prohibiting a transverse charge response are broken. Importantly, this highlights a fundamental distinction between the orbital and anomalous Hall responses: the latter relies on SOC and appropriate crystal symmetry, whereas the orbital Hall effect already captures the ferromagnet-like transport character of the altermagnet in the nonrelativistic limit in which altermagnetism itself emerges. Finally, we have extended this connection to odd-parity-wave magnets, where the p-wave spin splitting is accompanied by an orbital Hall effect with a non-antisymmetric conductivity tensor, reflecting the anisotropy of the underlying electronic structure. Our results thus identify orbital angular momentum as a natural link between crystal structure, unconventional magnetic order, and transverse transport.\\
\\
\\
\\
\\
\textbf{Methods}\\
\textbf{Spin and orbital angular momentum}\\
\noindent First, we diagonalize the Hamiltonian, as introduced in the main text, to determine the band structure as eigenenergies $E_{\nu\vec{k}}$. $\nu$ is the band index and $\vec{k}$ is the wave vector. Furthermore, we determine the eigenvectors $\ket{\nu\vec{k}}$ based on which we calculate the $z$ component of the spin texture
\begin{align}
    S_{z,\nu}(\vec{k})=\frac{\hbar}{2}\braopket{\nu\vec{k}}{\sigma_z}{\nu\vec{k}}.
\end{align}
Here, $\sigma_z$ is the corresponding Pauli matrix and $\hbar$ the reduced Planck constant. 
For the orbital angular momentum, we consider the modern theory accounting for off-diagonal elements~\cite{pezo2022orbital,gobel2024OHE}
\begin{align}
      &\braopket{\nu \vec{k}}{L_z}{\alpha \vec{k}} = \mathrm{i} \frac{e\hbar^2}{4g_L\mu_\mathrm{B}}  \sum_{\beta \neq \nu, \alpha} \left( \frac{1}{E_{\beta \vec{k}} - E_{\nu \vec{k}}} + \frac{1}{E_{\beta \vec{k}} - E_{\alpha \vec{k}}} \right)\notag\\
      &\times\left(\braopket{\nu \vec{k}}{v_x}{\beta \vec{k}} \braopket{\beta \vec{k}}{v_y}{\alpha \vec{k}} - \braopket{\nu \vec{k}}{v_y}{\beta \vec{k}} \braopket{\beta \vec{k}}{v_x}{\alpha \vec{k}}\right). \label{EQ:Lz_matrix_elements}
\end{align}
Here, $e$ is the electron charge, $g_L=1$ the Landé factor and $\mu_\mathrm{B}$ Bohr's magneton.
$L_{\nu,z}(\vec{k})=\braopket{\nu \vec{k}}{L_z}{\nu \vec{k}}$ are the diagonal elements and $\vec{v}=\frac{1}{\hbar}\nabla_{\vec{k}}H$ is the velocity operator. Since the orbital angular momentum scales with the lattice constant squared, we pick a lattice constant of $a=2.76\,$\AA\,  to be able to give comparable values for orbital and spin angular momentum. \\
\\
\textbf{Hall conductivities}\\
\noindent Based on these quantities we calculate the intrinsic Hall conductivities at zero temperature~\cite{nagaosa2010anomalous,pezo2022orbital,sinova2015spin}
\begin{align}
    \sigma_{i,j}(E_\text{F})&= -\frac{1}{2\pi}\frac{e^2}{h}\sum_\nu \int_{E_{\nu \vec{k}}\leq E_\text{F}}\Omega_{\nu,i,j}(\vec{k}) \,\mathrm{d}^2k\\
    \sigma^{L_z}_{i,j}(E_\text{F})&= \frac{1}{(2\pi)^2}\frac{e}{\hbar}\sum_\nu \int_{E_{\nu \vec{k}}\leq E_\text{F}}\Omega_{\nu,i,j}^{L_z}(\vec{k}) \,\mathrm{d}^2k,\\
    \sigma^{S_z}_{i,j}(E_\text{F})&= \frac{1}{(2\pi)^2}\frac{e}{\hbar}\sum_\nu \int_{E_{\nu \vec{k}}\leq E_\text{F}}\Omega_{\nu,i,j}^{S_z}(\vec{k}) \,\mathrm{d}^2k,
\end{align}
with $\{i,j\}=\{x,y,z\}$. The Hall conductivities are calculated as integrals of their respective Berry curvatures over all occupied states in the Brillouin zone~\cite{berry1984quantal}
\begin{align}
    \Omega_{\nu,i,j}(\vec{k})&= -2 \hbar^2\ \text{Im}\ \sum_{\mu\neq \nu} \frac{\braopket{\nu \vec{k}}{v_i}{\mu \vec{k}} \braopket{\mu \vec{k}}{v_j}{\nu \vec{k}}}{(E_{\nu \vec{k}} - E_{\mu \vec{k}})^2},\\
    \Omega_{\nu,i,j}^{L_z}(\vec{k})&= -2 \hbar^2\ \text{Im}\ \sum_{\mu\neq \nu} \frac{\braopket{\nu \vec{k}}{j_i^{L_z}}{\mu \vec{k}} \braopket{\mu \vec{k}}{v_j}{\nu \vec{k}}}{(E_{\nu \vec{k}} - E_{\mu \vec{k}})^2},\\
    \Omega_{\nu,i,j}^{S_z}(\vec{k})&= -2 \hbar^2\ \text{Im}\ \sum_{\mu\neq \nu} \frac{\braopket{\nu \vec{k}}{j_i^{S_z}}{\mu \vec{k}} \braopket{\mu \vec{k}}{v_j}{\nu \vec{k}}}{(E_{\nu \vec{k}} - E_{\mu \vec{k}})^2}.
\end{align}
For the orbital and spin Berry curvatures we use the respective current operators
\begin{align}
    j_i^{L_z}&=\frac{1}{2}(v_iL_z+L_zv_i),\\
    j_i^{S_z}&=\frac{1}{2}(v_iS_z+S_zv_i).
\end{align}\\
\\
\textbf{Data availability}\\
The data that supports the findings of this work is available from the authors on reasonable request.\\
\\
\textbf{Code availability}\\
The code that supports the findings of this work is available from the authors on reasonable request.\\
\\
\textbf{Acknowledgements}\\
This work was supported by the EIC Pathfinder OPEN grant 101129641 ``Orbital Engineering for Innovative Electronics'' and by Deutsche Forschungsgemeinschaft (DFG): Project No. 328545488 – CRC/TRR 227, Project No. B12 and by the German Excellence Strategy –EXC3112/1 –533767171 (Center for Chiral Electronics). The authors disclose the use of AI tools to assist with drafting and refining portions of the manuscript based on author-provided scientific ideas and context that were critically reviewed, verified, and revised.\\
\\
\textbf{Author contributions}\\
B.G. performed calculations, prepared the figures, planned the project and wrote the manuscript with significant inputs from E.S. and S.L.. B.G., E.S. and S.L. discussed the results.\\
\\
\textbf{Additional information}\\
\textbf{Supplementary information} The online version contains supplementary material available at [insert link].\\
\\
\textbf{Competing interests}\\
The authors declare no competing interests.\\
\\
\textbf{References}

\bibliography{short,MyLibrary}

@article{yahagi2024neel,
  title={Neel vector dependent orbital {H}all effect in altermagnetic {R}u{O}$_2$},
  author={Yahagi, Yuta},
  journal={arXiv preprint: 2410.01689},
  year={2024}
}

@article{zhang2026coexistence,
  title={Coexistence and Tunability of Orbital and Spin {H}all Effects in {R}u{O}$_2$},
  author={Zhang, Lishu and Zeer, Mahmoud and Go, Dongwook and Adamantopoulos, Theodoros and Schmitz, Peter and Bl{\"u}gel, Stefan and Niu, Chengwang and Mokrousov, Yuriy and Yan, Shishen and Yang, Hyunsoo and others},
  journal=PRL,
  volume={137},
  number={6},
  pages={066701},
  year={2026},
  publisher={APS},
  doi={10.1103/lf8d-26b6}
}

@article{goldberg1954new,
  title={New galvanomagnetic effect},
  author={Goldberg, Colman and Davis, RE},
  journal=PR,
  volume={94},
  number={5},
  pages={1121},
  year={1954},
  publisher={APS},
  doi={10.1103/PhysRev.94.1121}
}

@article{gobel2026exact,
  title={Exact theory of chirality-dependent p-wave magnetism and Edelstein effect in spin spirals},
  author={G{\"o}bel, B{\"o}rge and Saunderson, Tom G and Opfermann, Freia and Schimpf, Lennart and {\c{S}}a{\c{s}}{\i}o{\u{g}}lu, Ersoy and Lounis, Samir},
  journal={arXiv preprint: 2607.20094},
  year={2026}
}

@article{sasioglu2026chiral,
  title={Chiral-Angle-Controlled Altermagnetic Spin Splitting in Nanotubes},
  author={Sasioglu, Ersoy and Saunderson, Tom and G{\"o}bel, B{\"o}rge and Mertig, Ingrid and Lounis, Samir and others},
  journal={arXiv preprint: 2606.08757},
  year={2026}
}

@article{mazin2023induced,
  title={Induced Monolayer Altermagnetism in {M}n{P}({S},{S}e)$_3$ and {F}e{S}e},
  author={Mazin, Igor and Gonz{\'a}lez-Hern{\'a}ndez, Rafael and {\v{S}}mejkal, Libor},
  journal={arXiv preprint: 2309.02355},
  year={2023}
}

@article{jaeschke2025atomic,
  title={Atomic altermagnetism},
  author={Jaeschke-Ubiergo, Rodrigo and Bharadwaj, Venkata-Krishna and Campos, Warlley and Zarzuela, Ricardo and Biniskos, Nikolaos and Fernandes, Rafael M and Jungwirth, Tomas and Sinova, Jairo and {\v{S}}mejkal, Libor},
  journal={arXiv preprint: 2503.10797},
  year={2025}
}

@article{wei20252,
  title={{L}a$_2${O}$_3${M}n$_2${S}e$_2$: {A} correlated insulating layered d-wave altermagnet},
  author={Wei, Chao-Chun and Li, Xiaoyin and Hatt, Sabrina and Huai, Xudong and Liu, Jue and Singh, Birender and Kim, Kyung-Mo and Fernandes, Rafael M and Cardon, Paul and Zhao, Liuyan and others},
  journal={Physical Review Materials},
  volume={9},
  number={2},
  pages={024402},
  year={2025},
  publisher={APS},
  doi={10.1103/PhysRevMaterials.9.024402}
}

@article{parthenios2025spin,
  title={Spin and pair density waves in two-dimensional altermagnetic metals},
  author={Parthenios, Nikolaos and Bonetti, Pietro M and Gonz{\'a}lez-Hern{\'a}ndez, Rafael and Campos, Warlley H and {\v{S}}mejkal, Libor and Classen, Laura},
  journal=PRB,
  volume={112},
  number={21},
  pages={214410},
  year={2025},
  publisher={APS},
  doi={10.1103/llrq-1k9k}
}

@article{lieb1989two,
  title={Two theorems on the {H}ubbard model},
  author={Lieb, Elliott H},
  journal=PRL,
  volume={62},
  number={10},
  pages={1201},
  year={1989},
  publisher={APS},
  doi={10.1103/PhysRevLett.62.1201}
}

@article{chang2026inverse,
  title={Inverse {L}ieb materials: {A}ltermagnetism and more},
  author={Chang, Po-Hao and Belashchenko, Kirill D and Mazin, Igor I},
  journal={npj Quantum Materials},
  volume={11},
  pages={49},
  year={2026},
  publisher={Nature Publishing Group UK London},
  doi={10.1038/s41535-026-00880-w}
}

@article{brekke2023two,
  title={Two-dimensional altermagnets: Superconductivity in a minimal microscopic model},
  author={Brekke, Bj{\o}rnulf and Brataas, Arne and Sudb{\o}, Asle},
  journal=PRB,
  volume={108},
  number={22},
  pages={224421},
  year={2023},
  publisher={APS},
  doi={10.1103/PhysRevB.108.224421}
}

@article{jungwirth2025altermagnetism,
  title={Altermagnetism: {A}n unconventional spin-ordered phase of matter},
  author={Jungwirth, Tom{\'a}{\v{s}} and Fernandes, Rafael M and Fradkin, Eduardo and MacDonald, Allan H and Sinova, Jairo and {\v{S}}mejkal, Libor},
  journal={Newton},
  volume={1},
  number={6},
  year={2025},
  publisher={Elsevier},
  doi={10.1016/j.newton.2025.100162}
}

@article{song2025altermagnets,
  title={Altermagnets as a new class of functional materials},
  author={Song, Cheng and Bai, Hua and Zhou, Zhiyuan and Han, Lei and Reichlova, Helena and Dil, J Hugo and Liu, Junwei and Chen, Xianzhe and Pan, Feng},
  journal={Nature Reviews Materials},
  volume={10},
  number={6},
  pages={473--485},
  year={2025},
  publisher={Nature Publishing Group UK London},
  doi={10.1038/s41578-025-00779-1}
}

@article{hayami2019momentum,
  title={Momentum-dependent spin splitting by collinear antiferromagnetic ordering},
  author={Hayami, Satoru and Yanagi, Yuki and Kusunose, Hiroaki},
  journal={Journal of the Physical Society of Japan},
  volume={88},
  number={12},
  pages={123702},
  year={2019},
  publisher={The Physical Society of Japan},
  doi={10.7566/JPSJ.88.123702}
}

@article{bai2024altermagnetism,
  title={Altermagnetism: {E}xploring new frontiers in magnetism and spintronics},
  author={Bai, Ling and Feng, Wanxiang and Liu, Siyuan and {\v{S}}mejkal, Libor and Mokrousov, Yuriy and Yao, Yugui},
  journal={Advanced Functional Materials},
  volume={34},
  number={49},
  pages={2409327},
  year={2024},
  publisher={Wiley Online Library},
  doi={10.1002/adfm.202409327}
}

@article{tamang2025altermagnetism,
  title={Altermagnetism and altermagnets: {A} brief review},
  author={Tamang, Rupam and Gurung, Shivraj and Rai, Dibya Prakash and Brahimi, Samy and Lounis, Samir},
  journal={Magnetism},
  volume={5},
  number={3},
  pages={17},
  year={2025},
  publisher={MDPI},
  doi={10.3390/magnetism5030017}
}

@article{krempasky2024altermagnetic,
  title={Altermagnetic lifting of {K}ramers spin degeneracy},
  author={Krempask{\`y}, Juraj and {\v{S}}mejkal, L and D’souza, SW and Hajlaoui, M and Springholz, G and Uhl{\'\i}{\v{r}}ov{\'a}, K and Alarab, F and Constantinou, PC and Strocov, V and Usanov, D and others},
  journal=Nature,
  volume={626},
  number={7999},
  pages={517--522},
  year={2024},
  publisher={Nature Publishing Group UK London},
  doi={10.1038/s41586-023-06907-7}
}

@article{ma2021multifunctional,
  title={Multifunctional antiferromagnetic materials with giant piezomagnetism and noncollinear spin current},
  author={Ma, Hai-Yang and Hu, Mengli and Li, Nana and Liu, Jianpeng and Yao, Wang and Jia, Jin-Feng and Liu, Junwei},
  journal=NatureComm,
  volume={12},
  number={1},
  pages={2846},
  year={2021},
  publisher={Nature Publishing Group UK London},
  doi={10.1038/s41467-021-23127-7}
}

@article{hayami2020bottom,
  title={Bottom-up design of spin-split and reshaped electronic band structures in antiferromagnets without spin-orbit coupling: {P}rocedure on the basis of augmented multipoles},
  author={Hayami, Satoru and Yanagi, Yuki and Kusunose, Hiroaki},
  journal=PRB,
  volume={102},
  number={14},
  pages={144441},
  year={2020},
  publisher={APS},
  doi={10.1103/PhysRevB.102.144441}
}

@article{vsmejkal2020crystal,
  title={Crystal time-reversal symmetry breaking and spontaneous {H}all effect in collinear antiferromagnets},
  author={{\v{S}}mejkal, Libor and Gonz{\'a}lez-Hern{\'a}ndez, Rafael and Jungwirth, Tom{\'a}{\v{s}} and Sinova, Jairo},
  journal=SciAdv,
  volume={6},
  number={23},
  pages={eaaz8809},
  year={2020},
  publisher={American Association for the Advancement of Science},
  doi={10.1126/sciadv.aaz8809}
}

@article{yuan2021prediction,
  title={Prediction of low-{Z} collinear and noncollinear antiferromagnetic compounds having momentum-dependent spin splitting even without spin-orbit coupling},
  author={Yuan, Lin-Ding and Wang, Zhi and Luo, Jun-Wei and Zunger, Alex},
  journal=PRM,
  volume={5},
  number={1},
  pages={014409},
  year={2021},
  publisher={APS},
  doi={10.1103/PhysRevMaterials.5.014409}
}

@article{oh2026observation,
  title={Observation of spin-free interatomic orbital angular momentum in a chiral crystal},
  author={Oh, Dongjin and Hahn, Sungsoo and Pacella, Chiara and Yoo, Junseo and Rubio, Angel and Di Sante, Domenico and Kim, Changyoung},
  journal={arXiv preprint: 2605.21124},
  year={2026}
}

@article{brekke2024minimal,
  title={Minimal models and transport properties of unconventional p-wave magnets},
  author={Brekke, Bj{\o}rnulf and Sukhachov, Pavlo and Giil, Hans Gl{\o}ckner and Brataas, Arne and Linder, Jacob},
  journal=PRL,
  volume={133},
  number={23},
  pages={236703},
  year={2024},
  publisher={APS},
  doi={10.1103/PhysRevLett.133.236703}
}

@article{song2025electrical,
  title={Electrical switching of a p-wave magnet},
  author={Song, Qian and Stavri{\'c}, Srdjan and Barone, Paolo and Droghetti, Andrea and Antonenko, Daniil S and Venderbos, J{\"o}rn WF and Occhialini, Connor A and Ilyas, Batyr and Erge{\c{c}}en, Emre and Gedik, Nuh and others},
  journal=Nature,
  volume={642},
  number={8066},
  pages={64--70},
  year={2025},
  publisher={Nature Publishing Group UK London},
  doi={10.1038/s41586-025-09034-7}
}

@article{yamada2025metallic,
  title={A metallic p-wave magnet with commensurate spin helix},
  author={Yamada, Rinsuke and Birch, Max T and Baral, Priya R and Okumura, Shun and Nakano, Ryota and Gao, Shang and Ezawa, Motohiko and Nomoto, Takuya and Masell, Jan and Ishihara, Yuki and others},
  journal=Nature,
  volume={646},
  number={8086},
  pages={837--842},
  year={2025},
  publisher={Nature Publishing Group UK London},
  doi={10.1038/s41586-025-09633-4}
}

@article{hellenes2023p,
  title={P-wave magnets},
  author={Hellenes, Anna Birk and Jungwirth, Tom{\'a}{\v{s}} and Jaeschke-Ubiergo, Rodrigo and Chakraborty, Atasi and Sinova, Jairo and {\v{S}}mejkal, Libor},
  journal={arXiv preprint: 2309.01607},
  year={2023}
}

@article{vsmejkal2022emerging,
  title={Emerging research landscape of altermagnetism},
  author={{\v{S}}mejkal, Libor and Sinova, Jairo and Jungwirth, Tomas},
  journal=PRX,
  volume={12},
  number={4},
  pages={040501},
  year={2022},
  publisher={APS},
  doi={10.1103/PhysRevX.12.040501}
}

@article{fedchenko2024observation,
  title={Observation of time-reversal symmetry breaking in the band structure of altermagnetic {R}u{O}$_2$},
  author={Fedchenko, Olena and Min{\'a}r, Jan and Akashdeep, Akashdeep and D’souza, Sunil Wilfred and Vasilyev, Dmitry and Tkach, Olena and Odenbreit, Lukas and Nguyen, Quynh and Kutnyakhov, Dmytro and Wind, Nils and others},
  journal={Science Advances},
  volume={10},
  number={5},
  pages={eadj4883},
  year={2024},
  publisher={American Association for the Advancement of Science},
  doi={10.1126/sciadv.adj4883}
}

@article{smejkal2022beyond,
  author  = {Libor {\v{S}}mejkal and Jairo Sinova and Tom{\'a}{\v{s}} Jungwirth},
  title   = {Beyond Conventional Ferromagnetism and Antiferromagnetism: A Phase with Nonrelativistic Spin and Crystal Rotation Symmetry},
  journal = PRX,
  volume  = {12},
  pages   = {031042},
  year    = {2022},
  doi     = {10.1103/PhysRevX.12.031042}
}

@article{amin2024nanoscale,
  title={Nanoscale imaging and control of altermagnetism in {M}n{T}e},
  author={Amin, OJ and Dal Din, A and Golias, E and Niu, Y and Zakharov, A and Fromage, SC and Fields, CJB and Heywood, SL and Cousins, RB and Maccherozzi, F and others},
  journal=Nature,
  volume={636},
  number={8042},
  pages={348--353},
  year={2024},
  publisher={Nature Publishing Group UK London},
  doi={10.1038/s41586-024-08234-x}
}

@article{saunderson2026coupled,
  title={Coupled Spin-Orbital $ p $-Wave Magnetism via Structural and Magnetic Chirality},
  author={Saunderson, Tom G and G{\"o}bel, B{\"o}rge and {\c{S}}a{\c{s}}{\i}o{\u{g}}lu, Ersoy and Lounis, Samir},
  journal={arXiv preprint: 2607.02378},
  year={2026}
}

@article{gobel2025chirality2,
  title={Chirality-induced selectivity of angular momentum by orbital {E}delstein effect in carbon nanotubes},
  author={G{\"o}bel, B{\"o}rge and Mertig, Ingrid and Lounis, Samir},
  journal={Communications Physics},
  volume={8},
  number={1},
  pages={395},
  year={2025},
  publisher={Nature Publishing Group UK London},
  doi={10.1038/s42005-025-02331-7}
}

@article{gobel2025chirality,
  title={Chirality-induced orbital {E}delstein effect in an analytically solvable model},
  author={G{\"o}bel, B{\"o}rge and Schimpf, Lennart and Mertig, Ingrid},
  journal=PRR,
  volume={7},
  number={3},
  pages={033180},
  year={2025},
  doi={10.1103/vpjm-ntbh}
}

@article{yoda2015current,
  title={Current-induced orbital and spin magnetizations in crystals with helical structure},
  author={Yoda, Taiki and Yokoyama, Takehito and Murakami, Shuichi},
  journal={Scientific Reports},
  volume={5},
  number={1},
  pages={12024},
  year={2015},
  publisher={Nature Publishing Group UK London},
  doi={10.1038/srep12024}
}

@article{gobel2024topological,
  title={Topological orbital {H}all effect caused by skyrmions and antiferromagnetic skyrmions},
  author={G{\"o}bel, B{\"o}rge and Schimpf, Lennart and Mertig, Ingrid},
  journal={Communications Physics},
  volume={8},
  number={1},
  pages={17},
  year={2025},
  publisher={Nature Publishing Group UK London},
  doi={10.1038/s42005-024-01925-x}
}

@article{gobel2024OHE,
  title={{O}rbital {H}all {E}ffect {A}ccompanying {Q}uantum {H}all {E}ffect: {L}andau {L}evels {C}ause {O}rbital {P}olarized {E}dge {C}urrents},
  author={G\"obel, B\"orge and Mertig, Ingrid},
  journal=PRL,
  volume={133},
  pages={146301},
  year={2024},
  doi={10.1103/PhysRevLett.133.146301}
}

@article{lyalin2023magneto,
  title={Magneto-optical detection of the orbital {Ha}ll effect in chromium},
  author={Lyalin, Igor and Alikhah, Sanaz and Berritta, Marco and Oppeneer, Peter M and Kawakami, Roland K},
  journal=PRL,
  volume={131},
  number={15},
  pages={156702},
  year={2023},
  publisher={APS},
  doi={10.1103/PhysRevLett.131.156702}
}

@article{salemi2022theory,
  title={Theory of magnetic spin and orbital {H}all and {N}ernst effects in bulk ferromagnets},
  author={Salemi, Leandro and Oppeneer, Peter M},
  journal=PRB,
  volume={106},
  number={2},
  pages={024410},
  year={2022},
  publisher={APS},
  doi={10.1103/PhysRevB.106.024410}
}

@article{el2023observation,
  title={Observation of the orbital inverse {R}ashba--{E}delstein effect},
  author={El Hamdi, Anas and Chauleau, Jean-Yves and Boselli, Margherita and Thibault, Cl{\'e}mentine and Gorini, Cosimo and Smogunov, Alexander and Barreteau, Cyrille and Gariglio, Stefano and Triscone, Jean-Marc and Viret, Michel},
  journal=NaturePhys,
  volume={19},
  number={12},
  pages={1855--1860},
  year={2023},
  publisher={Nature Publishing Group UK London},
  doi={10.1038/s41567-023-02121-4}
}

@article{busch2023orbital,
  title={Orbital {H}all effect and orbital edge states caused by $s$ electrons},
  author={Busch, Oliver and Mertig, Ingrid and G{\"o}bel, B{\"o}rge},
  journal=PRR,
  volume={5},
  number={4},
  pages={043052},
  year={2023},
  publisher={APS},
  doi={10.1103/PhysRevResearch.5.043052}
}

@article{zhang2005intrinsic,
  title = {Intrinsic Spin and Orbital Angular Momentum {H}all Effect},
  author = {Zhang, S. and Yang, Z.},
  journal =PRL,
  volume = {94},
  issue = {6},
  pages = {066602},
  numpages = {4},
  year = {2005},
  month = {Feb},
  publisher = {American Physical Society},
  doi = {10.1103/PhysRevLett.94.066602},
  url = {https://link.aps.org/doi/10.1103/PhysRevLett.94.066602}
}

@article{bernevig2005orbitronics,
  title={Orbitronics: The intrinsic orbital current in $p$-doped silicon},
  author={Bernevig, B Andrei and Hughes, Taylor L and Zhang, Shou-Cheng},
  journal=PRL,
  volume={95},
  number={6},
  pages={066601},
  year={2005},
  publisher={APS},
  doi = {10.1103/PhysRevLett.95.066601},
  url = {https://link.aps.org/doi/10.1103/PhysRevLett.95.066601}
}

@article{kontani2008giant,
  title={Giant Intrinsic Spin and Orbital {H}all Effects in {Sr$_2M$O$_4$ ($M$= Ru, Rh, Mo)}},
  author={Kontani, Hiroshi and Tanaka, T and Hirashima, DS and Yamada, K and Inoue, J},
  journal=PRL,
  volume={100},
  number={9},
  pages={096601},
  year={2008},
  publisher={APS},
  doi = {10.1103/PhysRevLett.100.096601},
  url = {https://link.aps.org/doi/10.1103/PhysRevLett.100.096601}
}

@article{tanaka2008intrinsic,
  title={Intrinsic spin {H}all effect and orbital {H}all effect in {$4d$ and $5d$} transition metals},
  author={Tanaka, T and Kontani, Hiroshi and Naito, Masayuki and Naito, T and Hirashima, Dai S and Yamada, K and Inoue, J},
  journal=PRB,
  volume={77},
  number={16},
  pages={165117},
  year={2008},
  publisher={APS},
  doi = {10.1103/PhysRevB.77.165117},
  url = {https://link.aps.org/doi/10.1103/PhysRevB.77.165117}
}

@article{kontani2009giant,
  title={Giant orbital {H}all effect in transition metals: Origin of large spin and anomalous {H}all effects},
  author={Kontani, Hiroshi and Tanaka, T and Hirashima, DS and Yamada, K and Inoue, J},
  journal=PRL,
  volume={102},
  number={1},
  pages={016601},
  year={2009},
  publisher={APS},
  doi = {10.1103/PhysRevLett.102.016601},
  url = {https://link.aps.org/doi/10.1103/PhysRevLett.102.016601}
}

@article{go2018intrinsic,
  title={Intrinsic spin and orbital {H}all effects from orbital texture},
  author={Go, Dongwook and Jo, Daegeun and Kim, Changyoung and Lee, Hyun-Woo},
  journal=PRL,
  volume={121},
  number={8},
  pages={086602},
  year={2018},
  publisher={APS}, 
  doi = {10.1103/PhysRevLett.121.086602},
  url = {https://link.aps.org/doi/10.1103/PhysRevLett.121.086602}
}

@article{thonhauser2005orbital,
  title={Orbital magnetization in periodic insulators},
  author={Thonhauser, Timo and Ceresoli, Davide and Vanderbilt, David and Resta, Raffaele},
  journal=PRL,
  volume={95},
  number={13},
  pages={137205},
  year={2005},
  publisher={APS},
  doi = {10.1103/PhysRevLett.95.137205},
  url = {https://link.aps.org/doi/10.1103/PhysRevLett.95.137205}
}

@article{xiao2005berry,
  title={Berry phase correction to electron density of states in solids},
  author={Xiao, Di and Shi, Junren and Niu, Qian},
  journal=PRL,
  volume={95},
  number={13},
  pages={137204},
  year={2005},
  publisher={APS},
  doi = {10.1103/PhysRevLett.95.137204},
  url = {https://link.aps.org/doi/10.1103/PhysRevLett.95.137204}
}

@article{pezo2022orbital,
  title={Orbital {H}all effect in crystals: {I}nteratomic versus intra-atomic contributions},
  author={Pezo, Armando and Ovalle, Diego Garc{\'\i}a and Manchon, Aur{\'e}lien},
  journal=PRB,
  volume={106},
  number={10},
  pages={104414},
  year={2022},
  publisher={APS},
  doi = {10.1103/PhysRevB.106.104414},
  url = {https://link.aps.org/doi/10.1103/PhysRevB.106.104414}
}

@article{gobel2018magnetoelectric,
  title={Magnetoelectric effect and orbital magnetization in skyrmion crystals: {D}etection and characterization of skyrmions},
  author={G{\"o}bel, B{\"o}rge and Mook, Alexander and Henk, J{\"u}rgen and Mertig, Ingrid},
  journal=PRB,
  volume={99},
  number={6},
  pages={060406},
  year={2019},
  publisher={APS},
  doi={10.1103/PhysRevB.99.060406}
}

@article{chang1996berry,
  title={Berry phase, hyperorbits, and the {H}ofstadter spectrum: Semiclassical dynamics in magnetic {B}loch bands},
  author={Chang, Ming-Che and Niu, Qian},
  journal=PRB,
  volume={53},
  number={11},
  pages={7010},
  year={1996},
  publisher={APS},
  doi = {10.1103/PhysRevB.53.7010},
  url = {https://link.aps.org/doi/10.1103/PhysRevB.53.7010}
}

@article{ceresoli2006orbital,
  title={Orbital magnetization in crystalline solids: {M}ulti-band insulators, {C}hern insulators, and metals},
  author={Ceresoli, Davide and Thonhauser, Timo and Vanderbilt, David and Resta, Raffaele},
  journal=PRB,
  volume={74},
  number={2},
  pages={024408},
  year={2006},
  publisher={APS},
  doi={10.1103/PhysRevB.74.024408}
}

@article{raoux2015orbital,
  title={Orbital magnetism in coupled-bands models},
  author={Raoux, Arnaud and Pi{\'e}chon, Fr{\'e}d{\'e}ric and Fuchs, Jean-No{\"e}l and Montambaux, Gilles},
  journal=PRB,
  volume={91},
  number={8},
  pages={085120},
  year={2015},
  publisher={APS},
  doi={10.1103/PhysRevB.91.085120}
}

@article{kane2005z,
  title={{$Z_2$} topological order and the quantum spin {H}all effect},
  author={Kane, Charles L and Mele, Eugene J},
  journal=PRL,
  volume={95},
  number={14},
  pages={146802},
  year={2005},
  publisher={APS},
  doi={10.1103/PhysRevLett.95.146802}
}

@article{berry1984quantal,
  title={Quantal phase factors accompanying adiabatic changes},
  author={Berry, Michael V},
  journal={Proceedings of the Royal Society of London A: Mathematical, Physical and Engineering Sciences},
  volume={392},
  number={1802},
  pages={45--57},
  year={1984},
  organization={The Royal Society},
  doi={10.1098/rspa.1984.0023}
}

@article{nagaosa2010anomalous,
  title={Anomalous {H}all effect},
  author={Nagaosa, Naoto and Sinova, Jairo and Onoda, Shigeki and MacDonald, Allan H and Ong, Nai Phuan},
  journal=RevModPhys,
  volume={82},
  number={2},
  pages={1539},
  year={2010},
  publisher={American Physical Society},
  doi = {10.1103/RevModPhys.82.1539},
  url = {https://link.aps.org/doi/10.1103/RevModPhys.82.1539}
}

@article{sinova2015spin,
  title={Spin {H}all effects},
  author={Sinova, Jairo and Valenzuela, Sergio O and Wunderlich, J and Back, CH and Jungwirth, T},
  journal={Reviews of Modern Physics},
  volume={87},
  number={4},
  pages={1213},
  year={2015},
  publisher={APS},
  doi = {10.1103/RevModPhys.87.1213},
  url = {https://link.aps.org/doi/10.1103/RevModPhys.87.1213}
}

@article{go2017toward,
  title={Toward surface orbitronics: giant orbital magnetism from the orbital {R}ashba effect at the surface of $sp$-metals},
  author={Go, Dongwook and Hanke, Jan-Philipp and Buhl, Patrick M and Freimuth, Frank and Bihlmayer, Gustav and Lee, Hyun-Woo and Mokrousov, Yuriy and Bl{\"u}gel, Stefan},
  journal={Scientific Reports},
  volume={7},
  number={1},
  pages={46742},
  year={2017},
  publisher={Nature Publishing Group UK London},
  doi={10.1038/srep46742},
  url={10.1038/srep46742}
}

@Article{choi2023observation,
author={Choi, Young-Gwan
and Jo, Daegeun
and Ko, Kyung-Hun
and Go, Dongwook
and Kim, Kyung-Han
and Park, Hee Gyum
and Kim, Changyoung
and Min, Byoung-Chul
and Choi, Gyung-Min
and Lee, Hyun-Woo},
title={Observation of the orbital {H}all effect in a light metal {T}i},
journal={Nature},
year={2023},
month={Jul},
day={01},
volume={619},
number={7968},
pages={52-56},
issn={1476-4687},
doi={10.1038/s41586-023-06101-9},
url={10.1038/s41586-023-06101-9}
}

@String{Nature = "Nature"}

@String{NaturePhys = "Nature Phys."}

@String{NatureComm = "Nature Commun."}

@String{PR = "Phys.\ Rev."}

@String{PRB = "Phys.\ Rev.\ B"}

@String{PRM = "Phys.\ Rev.\ Mat."}

@String{PRX = "Phys.\ Rev.\ X"}

@String{PRL = "Phys.\ Rev.\ Lett."}

@String{PRR = "Phys.\ Rev.\ Res."}

@String{RevModPhys = "Rev.\ Mod.\ Phys."}

@String{SciAdv = "Science\ Adv."}



\end{document}